\documentstyle[aps,prl,twocolumn]{revtex}
 
\begin{document}

\wideabs{ 
\title{Superluminal Phenomena and the Quantum Preferred Frame}

\author{Jakub Rembieli{\'n}ski}

\address{Department of Theoretical Physics, University of {\L}{\'o}d{\'z},\\
  ul.~Pomorska 149/153, 90-236 {\L}{\'o}d{\'z}, Poland}

\date{\today}

\maketitle

\begin{abstract}
  Motivated by a number of recent experiments
  \cite{Stei1,Stei2,Nimi1,Nimi2,Wang1}, we discuss in this paper a
  speculative but physically admissible form and solutions of effective
  Maxwell-like equations describing propagation of electromagnetic
  field in a medium which ``feels'' a quantum preferred frame.
\end{abstract}

}

\paragraph{Introduction.}
 
In the last decade a number of experiments was performed where
propagation of some parameters describing the electromagnetic field
was found to be superluminal \cite{Stei1,Stei2,Nimi1,Nimi2,Wang1}.
This raises a serious interpretational problem of consistency with the
special relativity (SR) and Einstein causality if information is
associated with superluminally traveling wave.  Many authors
\cite{Chia1,Gari1} try to find reasonable interpretation of these
phenomena within framework of SR but it seems that the question still
remains open.  It is related to a delicate problem of the notion of
propagating signal and information \cite{Nimi2,Wang1,Chia1,Gari1}.  

On the other hand, in some experiments under considerations quantum
effects can play a crucial role.  For example, in \cite{Wang1} the
cesium atoms are promoted into excited quantum states and next
contribute to the light propagation.  So, analysis of this topics
should take a tension between SR and quantum mechanics (QM) into
account too.

As is well known, from an ``orthodox'' point of view there is a
``peaceful coexistence'' between SR and QM if a physical meaning is
attributed to final probabilities only \cite{Ahar1,Pere1,Pere2}.  However,
such restrictive approach is unsatisfactory for several physicists for
whom also the notion of a physical state, its time evolution,
localization of quantum events etc.\ should have a ``real'' not only
technical meaning.

According to this second line of understanding of QM we encounter a
number of theoretical problems on the borderline between QM and SR\@.
The most important ones are related to apparent nonlocality of QM and
lack of a manifest Lorentz covariance of quantum mechanics of systems
with finite degrees of freedom.  Recently several authors suggested
that a proper formulation of QM needs introduction of a preferred
frame (PF) \cite{Hard1,Perc1,Perc2,Remb2}.  In particular,
introduction of PF can solve some dilemmas related to causal
description of quantum collapse in the EPR-like experiments with
moving reference frames \cite{Suar1}.  It is important to stress that
the notion of PF used here is completely different from the
traditional one, related to ether and is in agreement with classical
experiments.  Most recent EPR experiments performed in Geneva
\cite{Gisi1} have been analyzed according to PF hypothesis and give a lower
bound for the speed of ``quantum information'' in the cosmic
background radiation frame (CBRF) at $2\times10^{4}c$.

A conceptual difficulty related to the PF notion lies in an apparent
contradiction with the Lorentz symmetry.  But as was shown in the
\cite{Remb2,Remb1,Remb3} this is not the case: it is possible to
arrange the Lorentz group transformation in such a way that the
Lorentz covariance survives while the relativity principle (democracy
between inertial frames) is broken.  Moreover such approach is
consistent with the classical phenomena.  Recall also that recently
some attention was devoted to the PF as a consequence of a possible
breaking of the Lorentz invariance \cite{Cole1,Coll1} in the high
energy physics.  We are not so ``radical'' in this paper because it is
enough to break the relativity principle only in order to extend the
causality notion and consequently to reconcile QM with the Lorentz
covariance.

We introduce and discuss a direct generalization of the macroscopic
(phenomenological) Maxwell equations which are both Lorentz-covariant
and ``feel'' the preferred frame.  We show that according to these
equations the electromagnetic field propagates faster than light in
the vacuum i.e.\ effective mass of photon is tachyonic.  Although our
derivation is purely classical it is motivated by the fact that in a
medium light propagation is mostly a quantum phenomenon; therefore the
influence of PF (if it really exists) can be in principle observed.
In the following we made simplified assumptions such as homogeneity
and isotropy of the medium.
        
\paragraph{The Lorentz covariance and the preferred frame.}

Because a ``folk theorem'' which states that local Lorentz covariance
implies relativity (i.e.\ democracy between inertial frames) is
commonly used, we begin with a brief review of the formalism
introduced in \cite{Remb2,Remb1,Remb3}.  Obviously, if we try to
realize the Lorentz group as a linear transformation of the Minkowski
coordinates only, the above mentioned ``theorem'' is necessarily true.
However, if a PF is distinguished we have in our disposal an
additional set of parameters, namely the four-velocity of PF with
respect to each inertial observer.  Using this freedom we can realize
the Lorentz group in a nonstandard way \cite{Remb1,Remb3}.
Physically, such a realization of the Lorentz transformations
corresponds to a nonstandard choice of the synchronization scheme for
clocks \cite{Ande1}.  In \cite{Remb2} this scheme was applied to
formulation of the manifestly covariant QM.

To be concrete, in that approach the Lorentz group is realized in a standard 
way as far as it is restricted to rotations, 
while for boosts we have
\begin{mathletters}
\label{eq:1}
\begin{eqnarray}
x'^{0} &=& \frac{1}{w^{0}} x^{0},\\
{\bf x'} &=& {\bf x} - {\bf w}\left( x^{0} + u^{0}({\bf u}\cdot{\bf x}) 
  - \frac{{\bf w}\cdot{\bf x}}{1+\sqrt{1+{\bf w}^{2}}} \right),
\end{eqnarray}
\end{mathletters}
and
\begin{mathletters}
\label{eq:2}
\begin{eqnarray}
u'^{0} &=& \frac{1}{w^{0}} u^{0},\\
{\bf u'} &=& {\bf u} - {\bf w}\left( \frac{1}{u^{0}} 
  - \frac{ {\bf u}\cdot{\bf w}}{1+\sqrt{1+{\bf w}^{2}}} \right),
\end{eqnarray}
\end{mathletters}
where $u^{\mu}=(u^{0},{\bf u})$ and $w^{\mu}=(w^{0},{\bf w})$ is the
(timelike) four-velocity of PF and $[x'^{\mu}]$, respectively as
observed from the inertial frame $[x^{\mu}]$.  The four-vectors
$u^{\mu}$ and $w^{\mu}$ are related to three-velocities via the
following formulae
\begin{mathletters}
\label{eq:3}
\begin{eqnarray}
{\bf v} &=& \frac{\bf u}{u^{0}},\\
{\bf V} &=& \frac{\bf w}{w^{0}},\\
\frac{1}{w^{0}} &=& \sqrt{ (1+u^{0}{\bf u}\cdot{\bf V} )^2 - {\bf V}^2 },
\end{eqnarray}
\end{mathletters}
The invariant line element $ds^{2}=g_{\mu\nu}(u)dx^{\mu}dx^{\nu}$, where
\begin{equation}
\left[g_{\mu\nu}\right]=
\left(\begin{array}{c|c}
1 & u^0 {\bf u}^T\\
\hline
u^0 {\bf u} & -I + (u^0)^2 {\bf u}\otimes {\bf u}^T 
\end{array}\right),
\end{equation}
has Minkowskian signature and it is easy to verify that the space line
element is $dl^{2}=d{\bf x}^{2}$.  The explicit relationship with the
standard (Einstein--Poincar\'{e}) synchronization is given by
$x^{0}_{E}=x^{0}+u^{0}{\bf u}\cdot{\bf x}$, ${\bf x}_{E}={\bf x}$, so
the time lapse in a space point is the same in both synchronizations.
Furthermore, the average light speed over closed loops is constant and
equal to the speed of light in vacuum (here $c=1$) in agreement with
the Michelson-like experiments.  It is important to stress that both
synchronizations (Einstein--Poincar\'{e} and the nonstandard one) lead
to the same results for velocities less or equal to the speed of light
but only the nonstandard synchronization scheme can be used for a
consistent description of possible superluminal phenomena
\cite{Remb3}.  This is because (as we see from
(\ref{eq:1})--(\ref{eq:2})) in the nonstandard synchronization the
Lorentz transformations have triangular form so the zeroth component
of a covariant four-vector is rescaled by a positive factor only.
Consequently, an absolute notion of causality can be introduced in
this framework.  Moreover, if superluminal propagation of information
do exist in reality, a PF {\it must be} distinguished and consequently
a convention of synchronization as well as the relativity principle
are broken.  An exhaustive discussion of the nonstandard formulation
of the Lorentz covariance in this language is done in
\cite{Remb2,Remb3}.

\paragraph{Effective Maxwell equations.}

In a homogeneous and isotropic medium the fields ${\bf D}$ and ${\bf
  H}$ are related to ${\bf E}$ and ${\bf B}$ via permittivity
$\varepsilon^{-1}$ and permeability $\mu$, respectively, where
$\varepsilon$ and $\mu$ are nonlinear functionals of ${\bf E}^{2}-{\bf
  B}^{2}$ and ${\bf E}\cdot{\bf B}$, in a nonlocal way.  To simplify
our considerations as far as possible, let us assume that in some
range of field intensity $\varepsilon$ and $\mu$ are slowly varying so
they can be treated approximately as constants.  Therefore, in our
equations we will use only ${\bf E}$ and ${\bf B}$ i.e.\ the
electromagnetic field tensor $F^{\mu\nu}$ and its dual
$\hat{F}^{\mu\nu}= \frac{1}{2} \varepsilon^{\mu\nu\sigma\lambda}
F_{\sigma\lambda}$.  Moreover, we assume that the possible (quantum)
response of the medium, related to preference by QM of a PF roughly
speaking is proportional to $\bf{E}$ and $\bf{B}$.  Under such
extremally simplified assumptions our phenomenological Maxwell-like
equations takes the form
\begin{mathletters}
\label{eq:5-6}
\begin{eqnarray}
\label{eq:5}
\partial_{\mu} F^{\mu\nu} +\alpha u_{\mu} F^{\mu\nu} &=& j^{\nu},\\
\label{eq:6}
\partial_{\mu} \hat{F}^{\mu\nu} +\beta u_{\mu} \hat{F}^{\mu\nu} &=& 0,
\end{eqnarray}
\end{mathletters} 
where $\alpha$ and $\beta$ are constants.  In the following we will
omit the induced current $j^{\nu}$ to concentrate on the consequences
of the influence of PF only.  It is not difficult to check that
equations (\ref{eq:5-6}) with $j^{\nu}=0$ have nontrivial solutions,
admitting a Fourier expansion, only for $\beta= -\alpha$, so
(\ref{eq:5-6}) must be replaced by \cite{foot1}
\begin{mathletters}
\label{eq:7-8}
\begin{eqnarray}
\label{eq:7}
\partial_{\mu} F^{\mu\nu}+ \alpha u_{\mu} F^{\mu\nu} &=& 0,\\
\label{eq:8}
\partial_{\mu} \hat{F}^{\mu\nu} -\alpha u_{\mu} \hat{F}^{\mu\nu} &=& 0,
\end{eqnarray}
\end{mathletters} 
with $\alpha$ depending on the properties and the state of the medium.
Of course, we can choose $\alpha\geq 0$.  Notice that (\ref{eq:8})
cannot be transformed to the form $\partial_{\mu} \hat{F}^{\mu\nu}=0$
by a duality transformation.  Obviously (\ref{eq:7-8}) are covariant
under transformations (\ref{eq:1})--(\ref{eq:2}).  Furthermore
(\ref{eq:7-8}) leads necessarily to the {\it tachyonic} wave equation
\begin{equation}
\label{eq:9}
\Box F^{\mu\nu}= \alpha^{2} F^{\mu\nu},
\end{equation}
where $\Box =g^{\mu\nu}(u)\partial_{\mu}\partial_{\nu}$.  In the
vacuum $\alpha_{\rm vac}=0$ (more precisely $\alpha_{\rm vac}<
2\times10^{-16}\;{\rm eV}$ \cite{Groo1}).  However in a ``PF feeling''
medium $\alpha$ should be different from zero.

As was shown in \cite{Remb3} Eq.~(\ref{eq:9}) can be consistently
quantized in the nonstandard synchronization and the resulting theory
is not plagued by pathologies related to quantization of tachyonic
field in the SR\@.  In particular in this framework vacuum is stable
\cite{Remb3}.  It is related to the fact that the spectral condition
$k^{0}>0$ is invariant also for space-like dispersion relation
$k^{2}<0$ (see transformation law (\ref{eq:1})).  A covariant
construction of the Fock space can be also done \cite{Remb3}.

It is easy to see that using (\ref{eq:8}) $F^{\mu\nu}$ can be
expressed by four-potential $A^{\mu}$ as
\begin{equation}
\label{eq:10}
F^{\mu\nu}=\partial^{\mu} A^{\nu}- \partial^{\nu} A^{\mu} 
-\alpha( u^{\mu} A^{\nu}- u^{\nu}A^{\mu} ),
\end{equation}
and the gauge transformations of $A^{\mu}$ are of the form $A^{\mu}
\to A^{\mu}+( \partial^{\mu}- \alpha u^{\mu} )\chi$.
Therefore, the above field equations can be derived from the
Lagrangian density
\begin{eqnarray}
L&=&-\frac{1}{4} F_{\mu\nu} F^{\mu\nu}+\frac{1}{2}F_{\mu\nu}
\left[\partial^{\mu}A^{\nu}-\partial^{\nu}A^{\mu}\right.\nonumber\\
&&\left.-\alpha (u^{\mu}A^{\nu}-u^{\nu}A^{\mu})\right].
\label{eq:11}
\end{eqnarray}
For a general field $F^{\mu\nu}$ and under standard identification of
$F^{\mu\nu}$ with ${\bf E}$ and ${\bf B}$ ($F^{0k}=E^{k}$,
$F^{ij}=\varepsilon^{ijk}B^{k}$) the Lorentz invariants $F\hat{F}$ and
$F^{2}$ are
\begin{mathletters}
\label{eq:12}
\begin{eqnarray} 
F^{\mu\nu}\hat{F}_{\mu\nu} &=& -4{\bf E}\cdot{\bf B},\\
F^{\mu\nu} F_{\mu\nu} &=& -{\rm Tr}(gFgF)\nonumber\\ 
&=& 2({\bf B}^{2} -{\bf E}^{2}) +4u^{0}{\bf u}\cdot({\bf B}\times{\bf E})
\nonumber\\
&&-2(u^0)^2({\bf u}\times{\bf B})^{2}.
\end{eqnarray} 
\end{mathletters}
Now let us examine the monochromatic plane wave solutions $f^{\mu\nu}$
of (\ref{eq:7-8}).  Let
\begin{equation}
\label{eq:13}
f^{\mu\nu}=e^{\mu\nu}(k)e^{ikx} + e^{*\mu\nu}(k)e^{-ikx},
\end{equation} 
where $kx=k_{\mu}x^{\mu}$. Therefore, by (\ref{eq:7-8}) we find
\begin{mathletters}
\label{eq:14}
\begin{eqnarray}
( ik_{\mu}+ \alpha u_{\mu} )e^{\mu\nu} &=& 0,\\
( ik_{\mu}- \alpha u_{\mu} ){\hat{e}}^{\mu\nu} &=& 0,
\end{eqnarray}
\end{mathletters} 
and (\ref{eq:14}) lead to the tachyonic dispersion relation
$k^{2}=-\alpha^{2}$.  The solution of the system (\ref{eq:14}) has the
form
\begin{eqnarray}
e^{\mu\nu}&=&\left(\frac{\alpha (un)+ i(kn)}{\alpha 
    + i(uk)}\right)(k^{\mu}u^{\nu}-k^{\nu}u^{\mu})\nonumber\\
&&- (k^{\mu}n^{\nu}-k^{\nu}n^{\mu})
- i \alpha (u^{\mu}n^{\nu}-u^{\nu}n^{\mu}),
\label{eq:15}
\end{eqnarray} 
where $k^{\mu}$,$u^{\mu}$,$n^{\mu}$ and
$\varepsilon^{\mu\nu\sigma\lambda} k_{\nu}u_{\sigma}n_{\lambda}$ span
a basis, $un=u_{\mu} n^{\mu}$ etc.\ and $n^{\mu}$ can be complex in
general.

It is convenient to consider our plane wave solution in the preferred
frame.  If PF is realized as the cosmic background radiation frame,
this choice is reasonable from our point of view because $v_{\rm
  solar}\approx369.3\pm2.5\;{\rm km}/{\rm s} \ll c$ with respect to
CBRF \cite{Line1}.  For PF, $u^{\mu}=(1,{\bf 0})$ so in this frame
$g_{\mu\nu}={\rm diag}(1,-1,-1,-1)$.  Now we can put ${\bf n}=-({\bf
  a} + i {\bf b}) e^{i\varphi} /2$, where $\bf{a}$ and $\bf{b}$ are
real and ${\bf a} \perp {\bf b}$.  Thus from (\ref{eq:15}) we have the
following form of the electromagnetic wave in the preferred frame
\begin{mathletters}
\label{eq:16}
\begin{eqnarray}
{\bf E} &=& \frac{1}{|\bf k|}{\bf k}\times \{{\bf k}\times 
[-\cos{(kx+\varphi+\xi)}{\bf a}\nonumber\\
&& + \sin{(kx+\varphi+\xi)}{\bf b}]\},\\
{\bf B} &=& {\bf k}\times [\cos{(kx+\varphi)}{\bf a}
-\sin{(kx+\varphi)}{\bf b}]
\end{eqnarray}
\end{mathletters}
where $\xi=\arccos{(k^{0}/|{\bf k}|)}$, $|{\bf k}| >\alpha$,
$k^{0}=\sqrt{|{\bf k}|^{2}-\alpha^2}$.  Evidently, we can choose ${\bf
  a} \perp {\bf k}$ and ${\bf b}\perp {\bf k}$.  Therefore in the PF
\begin{equation}
\label{eq:17}
-\frac{1}{4} F\hat{F}={\bf E}\cdot{\bf B}=\pm\alpha 
|{\bf a}||{\bf b}||{\bf k}|
\end{equation}
and
\begin{eqnarray}
\frac{1}{2} F^{2}&=&{\bf B}^{2}-{\bf E}^2\nonumber\\
&=&\alpha ({\bf a}^2-{\bf b}^2)|{\bf k}|\sin{(2kx+2\varphi+\xi)} 
\label{eq:18}
\end{eqnarray}
Therefore, contrary to the massless case, $F\hat{F}$ and $F^{2}$
cannot vanish simultaneously except the case ${\bf E}={\bf B}=0$.
However, both ${\bf E}$ and ${\bf B}$ are perpendicular to ${\bf k}$
so the wave front propagates along ${\bf k}$.  Moreover, the angle
between ${\bf E}$ and ${\bf B}$ is constant in time.  The linear
polarization is obtained for ${\bf a}=0$ or ${\bf b}=0$; in this case
${\bf E}\perp{\bf B}$. The elliptic polarization is obtained for ${\bf
  a}$ and ${\bf b}$ simultaneously different from zero; in this case
${\bf E}\cdot{\bf B}\neq 0$.  Notice that for $\alpha$ going to zero
we obtain standard vacuum solution.

Now, the group velocity of the electromagnetic wave (\ref{eq:16}) is
superluminal
\begin{equation}
\label{eq:19}
{\bf v}_g=\nabla_{k}\omega({\bf k})=\left(\frac{\sqrt{k^{02}
      +\alpha^{2}}}{k^{0}}\right)\frac{\bf k}{|{\bf k}|}
\end{equation}
while phase propagates subluminally
\begin{equation}
\label{eq:20}
{\bf v}_{\rm ph}=\left(\frac{k^{0}}{\sqrt{k^{02}
      +\alpha^{2}}}\right)\frac{\bf k}{|{\bf k}|}.
\end{equation}     

A very important question is the energy transport associated with the
electromagnetic wave.  The locally conserved canonical energy-momentum
tensor, derived from the Lagrangian (\ref{eq:11}), is of the form
\begin{equation}
\label{eq:21}
T_{\mu}^{\nu}=\frac{1}{4}\delta_{\mu}^{\nu} F^{2}-F_{\nu\lambda} 
F^{\mu\lambda}-\alpha F_{\nu\lambda} A^{\lambda} u^{\mu}.
\end{equation}
It is evidently neither gauge-invariant nor is $T_{\mu}^{\nu}$
symmetrical in $\mu$ and $\nu$.  While this second deficiency is not
serious, the first one is very unpleasant and the question how to
remedy this problem is unclear because the standard procedure fails in
this case.  However the field four-momentum
\begin{equation}
\label{eq:22}
P_{\mu}:=\int_{t=\rm const} d\sigma^{\nu} T_{\nu\mu}
=\int d^{3}{\bf x} T_{0\mu} 
\end{equation}
is gauge-invariant.  Notice that the transformation law (\ref{eq:1})
implies that the invariant-time hyperplane $x^{0}={\rm const}$ (i.e.\ 
$dx^{0}=0$) is an invariant notion in our framework. Therefore we can
express the volume element $d^{3}{\bf x}$ via the obvious relation
\begin{equation}
\label{eq:23}
u^{0} d^{3}{\bf x}=-u_{\mu}d\sigma^{\mu}\equiv -ud\sigma
\end{equation}
which holds because in all inertial frames space components of
$u_{\mu}=g_{\mu\nu}(u) u^{\nu}$ are equal to zero i.e.\ $u_{k}=0$ for
$k=1, 2, 3$. Thus the volume
\begin{equation}
\label{eq:24}
V=-\frac{1}{u^{0}} \int_{t=\rm const} u d\sigma 
\end{equation}
transforms under the Lorentz transformations
(\ref{eq:1})--(\ref{eq:2}) according to the law
\begin{equation}
\label{eq:25}
V'=w^{0}V. 
\end{equation}
This fact enables us to define the covariant four-momentum per volume
as well as the gauge-invariant average density
\begin{equation}
\label{eq:26}
p^{\mu}:=\lim_{V \to 0} \frac{1}{V} \int_{V} d^{3} {\bf x} 
T_{0}^{\mu} 
\end{equation}
Now, for the monochromatic plane wave (\ref{eq:16}) in the PF,
Eq.~(\ref{eq:26}) leads to
\begin{mathletters}
\label{eq:27}
\begin{eqnarray}
p^{0} &=& \frac{(k^0)^2}{2} ({\bf a}^2+{\bf b}^2),\\
{\bf p} &=& \frac{{\bf k} k^0}{2}({\bf a}^2+{\bf b}^2).
\end{eqnarray}
\end{mathletters}
Thus, in the PF
\begin{equation}
\label{eq:28}
(p^0)^2-{\bf p}^{2}=-\alpha^{2} (k^0)^2\frac{({\bf a}^2+{\bf b}^2)^{2}}{4}
\leq 0 
\end{equation}
i.e.\ the energy transport is superluminal too in this case.  Of
course, the statements resulting from (\ref{eq:19}), (\ref{eq:20}) and
(\ref{eq:28}) are true in all inertial frames by the Lorentz
covariance.

Finally, wave packets are obtained by use of the invariant measure
\cite{Remb3}
\begin{equation}
\label{eq:29}
d\mu (k,\alpha)=\theta (k^0)\delta (k^2 +\alpha^2)d^{4}k,
\end{equation}   
which selects covariantly only the upper part of the momentum
hyperboloid and determines the range of integration over
$k_{i}$, $i=1,2,3$.  Namely
\begin{equation}
\label{eq:30}
F^{\mu\nu}=\int d\mu (k,\alpha) f^{\mu\nu}(k,u,n(k)).
\end{equation}     

\paragraph{Conclusions.}

Our discussion shows that a possible influence of the quantum
preferred frame on an appropriate medium can cause tachyoniclike
propagation for the electromagnetic waves.  It is interesting that
solutions for the effective Maxwell equations (\ref{eq:7-8}) are very
regular and similar to the usual ones. Therefore, it seems that this
model offers an alternative for standard proposals of explaining of
the superluminal phenomena.

\paragraph*{Acknowledgment.}

I acknowledge discussion with Piotr Kosi{\'n}ski and Wac{\l}aw Tybor.
This paper was financially supported by the University of Lodz.


\begin{thebibliography}{99}
\bibitem{foot1} This can be done equivalently by an appropriate
  redefinition $F^{\mu\nu}\to\exp(\gamma ux)F^{\mu\nu}$.
\end{thebibliography}
\end{document}